\documentstyle[prl,aps,psfig,twocolumn]{revtex}
\tighten
\def\dens{\langle n \rangle}
\def\etal{{\it et al.~}\ }           

\begin{document}

\draft

\title{Evolution of the Density of States Gap in 
a Disordered Superconductor}

\author{Carey Huscroft and Richard T. Scalettar}

\address{Department of Physics, University of California,
Davis, CA 95616}

\date{\today}

\address{%
\begin{minipage}[t]{6.0in}
\begin{abstract}
It has only recently been possible to study the 
superconducting state in the attractive Hubbard Hamiltonian 
via a direct observation of the formation of a gap 
in the density of states $N(\omega)$.  Here we determine the effect of
random chemical potentials on $N(\omega)$ and show
that at weak coupling, disorder 
closes the gap concurrently with the destruction of
superconductivity.  At larger, but still intermediate coupling,
a pseudo--gap in $N(\omega)$
remains even well beyond the point at which off--diagonal
long range order vanishes.
This change in the elementary excitations of the insulating phase
corresponds to a crossover between Fermi-- and Bose--Insulators.
These calculations represent the first computation 
of the density of states in
a finite dimensional disordered fermion model via the
Quantum Monte Carlo and maximum entropy methods.
\typeout{polish abstract}
\end{abstract}
\end{minipage}}

\maketitle
\narrowtext

Analysis of the effect of randomness on the superconducting state
has a long history,\cite{PLEE} 
beginning with Anderson's observation\cite{ANDERSON}
that in the presence of non-magnetic disorder the 
Cooper wavefunction can be constructed by 
replacing $(k \uparrow, -k \downarrow)$ pairs by an appropriate
combination of time reversed, but still extended, eigenstates.
Recent theoretical work\cite{UNIVCONDT} has attempted to
understand experiments\cite{UNIVCONDE} on the superconducting--insulator
(SC--I) transition in thin films. 
Here, the superconducting phase is destroyed by strong disorder which
localizes the electrons entirely.
Numerical work\cite{UNIVCONDN} has focussed on 
boson models in which coherence of the pair phase is the central
issue, and the role of fluctuations in the pair amplitude is
suppressed.   Granular systems might be well described by 
such pre--formed Cooper pairs, and, furthermore, 
scaling arguments suggest universal conductivity might
find a natural explanation within boson systems.\cite{UNIVCONDT}

Despite the successes of numeric studies of interacting,
disordered bose systems, it is clear that explicit calculations for 
itinerant electron models, in which pairs can break apart, are essential.
Experimentally, there is evidence for both 
SC--I transitions to a Bosonic state\cite{PAALANEN}
and to a Fermionic state\cite{VALLES}, 
distinguished by the existence
or lack of an energy gap, respectively, in the insulating phase.
There is currently considerable controversy regarding the meaning of
the two insulting phases seen experimentally.\cite{DAS}  
Theories implicating both Bosons\cite{FISHER} 
and Fermions\cite{BELITZ2} exist.

In this paper we describe the evolution of the 
single-particle density of states
for the disordered, attractive Hubbard Hamiltonian, 
a model which can interpolate between Bose and Fermi limits.  
We show that increasing disorder destroys long
range pairing correlations and drives a superconducting--insulator phase
transition.  However, the density of states $N(\omega)$
shows a gap which closes with the destruction of superconducting 
long-range order for relatively weak couplings yet
retains a gap beyond the critical disorder strength for larger couplings.
Sweeps across the disorder--interaction phase diagram allow us to locate
quantitatively the critical coupling strengths for gap and pair
order formation in a set of
representative cases.  

The attractive Hubbard Hamiltonian, in the presence of random site 
energies is,
\begin{eqnarray}
H = &-&t\sum_{\langle {\bf ij} \rangle \sigma} 
( c_{{\bf i} \sigma}^{\dagger}c_{{\bf j} \sigma}
+ c_{{\bf j} \sigma}^{\dagger}c_{{\bf i} \sigma} ) 
 - \sum_{{\bf i} \sigma} (\mu - v_{{\bf i}}) (n_{{\bf i} \sigma} -\frac12)
\nonumber \\
&-& |U| \sum_{{\bf i}} (n_{{\bf i} \uparrow}-\frac12)
( n_{{\bf i} \downarrow}-\frac12).
\label {eq:eq1}
\end{eqnarray}
Here $c_{{\bf i}\sigma}$ is a fermion destruction operator at site ${\bf i}$
with spin $\sigma$, 
$n_{{\bf i}\sigma}=c_{{\bf i}\sigma}^{\dagger}c_{{\bf i}\sigma}$,
and the chemical potential $\mu$ fixes the average density $\dens$. 
The site energies $v_{{\bf i}}$ are independent random variables 
with a uniform distribution over $[-V/2,V/2]$.  
The interaction has been written in particle--hole
symmetric form so that $\mu=0$ corresponds to $\dens=1$ at
all $U$ and $T$ when $V=0$.       
The lattice sum $\langle {\bf ij} \rangle$ is over
near neighbor sites on a two dimensional square lattice.       

We solve for the equilibrium properties of this Hamiltonian
using the determinant QMC method.\cite{SUGAR}
Previous numerical studies of the clean model have determined
the phase diagram\cite{NEGUSIMPD} by a finite size
scaling analysis of the equal time pair--pair correlation functions,
\begin{equation}
P_s ({\bf j}) = {1 \over N} \sum_{\bf i} \langle \,\, 
\Delta_{{\bf i}+{\bf j}}^{ \,}
\Delta_{{\bf i}}^{\dagger} \,\, \rangle,
\hskip0.3in 
\Delta_{{\bf j}}^{\dagger}  = c_{\uparrow {\bf j}}^{\dagger}
c_{\downarrow {\bf j}}^{\dagger}.
\label {eq:eq2}
\end{equation}                
The phase diagram consists of a state with simultaneous charge density
wave (CDW) and superconducting correlations at $T=0$ and half--filling,
and a finite temperature Kosterlitz--Thouless transition
(with a maximal $T_{c} \approx 0.1 t$ for $|U|=4t$)
to a phase with power law decay of pairing correlations
off half--filling.
In this paper we will study fillings $\dens=0.875$ for which
strong CDW correlations are absent and the transition
temperature to the SC phase is nearly maximal.

In Fig.~1 we show the suppression of long range 
correlations in $P_{s}({\bf j})$ with increasing disorder strength.
A finite size scaling analysis of the structure factor
determines the critical value 
$V_{c} = (3.25 \pm 0.2) t$ 
for the destruction of the superconducting state.  $V_{c}$
can also be identified by the superfluid
density $D_{s}$ and the conductivity.\cite{NEGUUC}
These quantities give consistent values, 
$V_{c} = (3.2 \pm 0.7) t$ and
$V_{c} = (3.5 \pm 0.5) t$, respectively,
for the critical disorder.

\begin{figure}
\hspace*{-0.2in}
\psfig{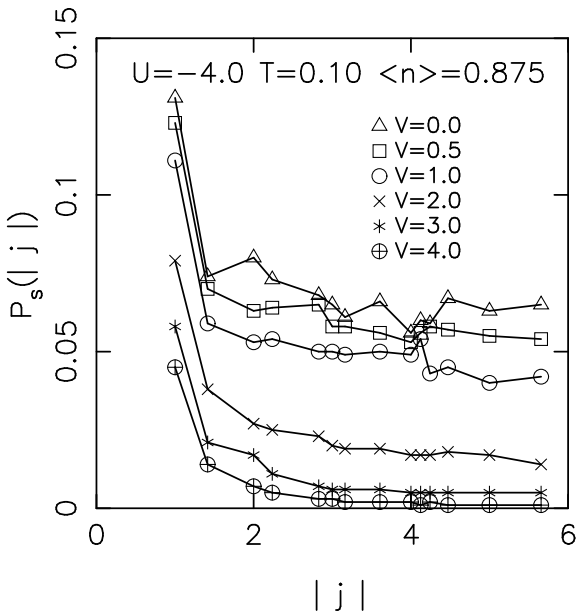}
\bigskip
\caption{             
The equal time pair correlation $P_s({\bf j})$
is shown at $T=0.10$ as a function of separation ${\bf j}$
for different degrees of disorder.
$P_s({\bf j})$
remains finite at large ${\bf j}$ for $V=2.0$, but
goes to zero rapidly with ${\bf j}$ at $V=4.0.$
}
\end{figure}

The density of states is a quantity which, like the conductivity,
is directly accessible experimentally. Its evaluation
requires an inversion of the integral relation,
\begin{equation}
G(\tau) = 
{1 \over N} \sum_{{\bf p}}
\langle c_{{\bf p}}(\tau)c^{\dagger}_{{\bf p}}(0) \rangle=
\int_{-\infty}^{+\infty} d\omega \,\,
{e^{-\omega \tau} N(\omega) \over 1 + e^{-\beta \omega}},
\label {eq:eq3}
\end{equation}
between the density of states $N(\omega)$ and the imaginary time dependent
Greens function $G(\tau)$.
We do this using 
Bryan's method\cite{BRYAN} and Classic Maximum
Entropy\cite{GUBERNATISandJARRELL}, using the full
imaginary--time covariance matrix.
                                             
Although this approach is fairly well--developed for clean systems,
the inclusion of disorder raises a number of new questions of principle.  
The proper treatment of errors and correlations in the
QMC data was the central achievement of the ME
approach,\cite{ERRORS} and it is not obvious how
additional fluctuations from disorder averaging
will affect the procedure.  Therefore, we have checked
our calculation against analytic results
both at weak and strong coupling.  Besides
verifying the numerics, the behavior in these two limits 
also presages the two distinct types of insulating behavior
we see in the full model.

In the non--interacting limit
we can diagonalize
the Hamiltonian and obtain $E_{\alpha}$, the single particle
eigenenergies and thence
$N(\omega)={1 \over N} \sum_{\alpha} \delta(\omega - E_{\alpha})$.
In this limit the determinant
QMC method computes the imaginary time fermion
Greens function exactly.  Tests of 
the ME approach in the clean case
require that some noise be added to the exact $G(\tau)$ coming from the 
QMC.\cite{WHITE}  In the random case, this {\it ad hoc} noise does not need to
be added, since $G(\tau)$ already has error bars from
disorder averaging.  We find that the ME approach exactly tracks the disorder
induced broadening of $N(\omega)$.\cite{USUNPUB}
For small $|U|$ we expect disorder to broaden and
eliminate the superconducting gap as well.

The strong coupling, $t=0$, limit also can be solved analytically.
For a single site,
\begin{eqnarray}
G(\tau)   &=&  { 
e^{-\beta U /4} e^{-\tau \Delta_{-}} 
+ e^{\beta(U/4 + v_{{\bf i}})} e^{-\tau \Delta_{+}} \over
2 (e^{\beta U / 4} {\rm cosh} \beta v_{{\bf i}} + e^{-\beta U / 4}) },
\nonumber\\
N(\omega) &=&  
{ (1+e^{-\beta \Delta_{-}}) e^{-\beta U /4} \over
2 (e^{\beta U / 4} {\rm cosh} \beta v_{{\bf i}} + e^{-\beta U / 4}) }
\delta (\omega-\Delta_{-})
\nonumber\\
&+& 
{ (1+e^{-\beta \Delta_{+}}) e^{\beta(U/4 + v_{{\bf i}}) } \over
2 (e^{\beta U / 4} {\rm cosh} \beta v_{{\bf i}} + e^{-\beta U / 4}) }
\delta (\omega-\Delta_{+}).
\label {eq:eq4}
\end{eqnarray}
Here $\Delta_{\pm}=\pm U/2 + v_{{\bf i}}$. 
After disorder averaging, the two delta functions in the
density of states $N(\omega)$ are broadened to two distributions
centered about $\pm U/2$, each with width $V$.  When $V=U$
these merge, and the gap in $N(\omega)$ at $\omega=0$ is closed.
However, the thermal factors greatly suppress $N(\omega)$
near $\omega=0$.
At $t=0$ there is no long range pairing
order, so that this ``pseudo--gap'' in the density of states
reflects the tendency for on--site singlet formation.
We have shown that our QMC+ME procedure accurately reproduces 
the analytic result
Eq.~4 as a function of temperature, disorder strength, and interaction 
strength.\cite{USUNPUB}  A central conclusion of our work is that
this pseudogap behavior persists far from the strong--coupling limit
as the hopping $t$ is turned on.

We now study the cross--over between these two possible types of
effect of disorder on $N(\omega)$ in the full model.
In Fig.~2 we show the evolution of $N(\omega)$ with increasing
interaction strength $|U|$ at fixed disorder $V=2t$.  $N(\omega)$
evolves from its gapless, noninteracting form to possessing
a well--formed gap at $|U|=4t$.  Similarly, 
in Fig.~3 we show the evolution of $N(\omega)$ with increasing disorder
$V$, at fixed values of the interaction, $|U|=4t$.
In contrast to the non--interacting case where
increasing $V$ results in the expected broadening of $N(\omega)$,
here the gap in $N(\omega)$ is remarkably robust.  Note the values of
$V$ are well beyond the point where the superconducting--insulator
transition has occurred, as indicated by the results in Fig.~1
and related measurements.\cite{NEGUUC}
In the repulsive Hubbard model at the same value
of $U/t$ the gap in $N(\omega)$ 
requires {\it long range} antiferromagnetic order be present,
specifically, the correlation length $\xi_{{\rm af}}$
must exceed the lattice size.\cite{WHITE2}
In contrast, here we find that a gap in $N(\omega)$ persists despite
the clear destruction of such long range order.

\begin{figure}
\hspace*{-0.2in}
\psfig{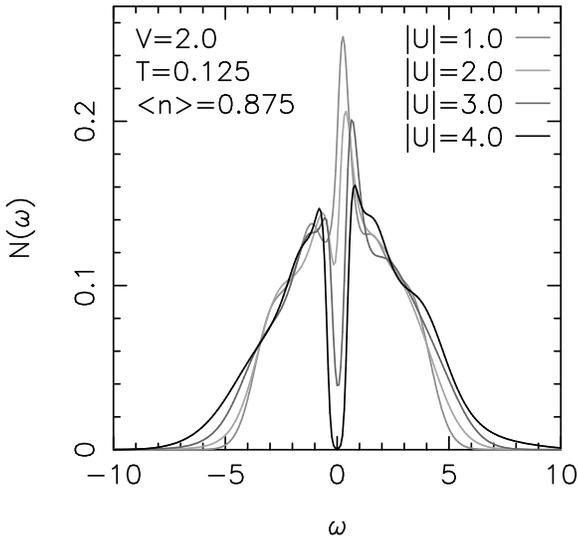}
\bigskip
\caption{             
The density of states $N(\omega)$ at $t=1$, $T=0.125$, $V=2$, $|U|=1,2,3,4$
on an $8 \times 8$ lattice.  For sufficiently small $|U|$,
scattering by the disorder inhibits phase coherence and the
superconducting gap is destroyed.
A superconducting
gap forms with increasing interaction strength $U$.  A study of the
pair correlations $P_s({\bf j})$ indicates that gap formation
occurs precisely with the onset of superconductivity.
}
\end{figure}

How can we interpret these results?  In the weak--coupling limit, 
disorder broadens $N(\omega)$ at all energies.  In particular,
the superconducting gap in $N(\omega)$
is closed at the same point where 
disorder destroys the long--range pair correlations.
This closing of the gap is indicative
of the predominance of fermionic elementary excitations in the insulating
phase.  In the strong coupling limit,
the density of states is not broadened by disorder near $\omega=0$ and
a pseudogap survives.
The results of Fig.~3 imply
that already at a value of interaction strength $|U|=4t$
which is only one--half the bandwidth, $W=8t$, the superconducting
gap is robust to disorder.  In previous work,\cite{NEGUSIMSG} it was found that
at $V=0$ and $|U|=4t$ the spin susceptibility exhibits a reduction at low
temperatures, indicative of bose--like behavior, despite the 
fact that the temperature dependence of the chemical potential 
$\mu(T)$ is still clearly that of a degenerate Fermi system.
We conclude here that 
for an interaction strength as low as $|U|=4t$,
the density of states reflects bose--like character at all $V$.  
As shown in Fig.~3, at either side of the critical disorder $V_c$ for
the destruction of long--range order in this coupling regime, there are
no indications of fermionic excitations in $N(\omega)$.  

\begin{figure}
\hspace*{-0.2in}
\psfig{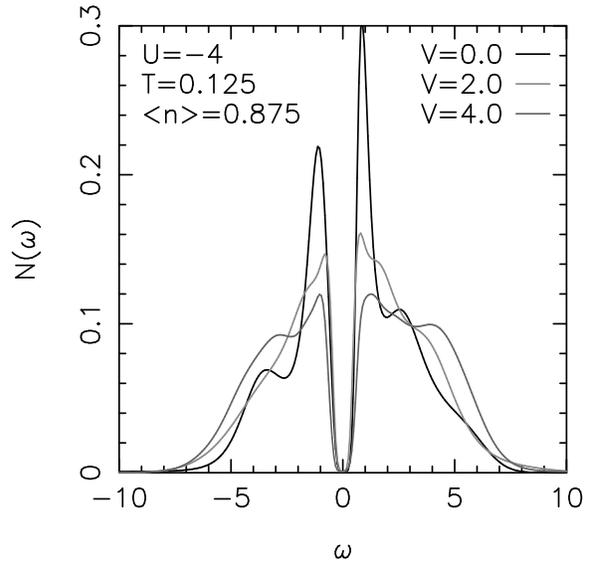}
\bigskip
\caption{             
The density of states $N(\omega)$ at $t=1$, $U=-4$, $T=0.125$,
$V=0.0,2.0,4.0$ on an
$8 \times 8$ lattice.  A study of the equal time pair correlation shows
that for $V < V_c \approx 3.5$ 
the system is a superconductor.  For $V \geq V_c$
the system is an insulator.  The gap in $N(\omega)$ persists even
after long-range pair correlations, and superconductivity, are destroyed.
Furthermore, there is no enhancement in $N(\omega)$ as one approaches
$V_c$ from either side, indicating that the system is not
a Fermi liquid at the SC--BI transition.
}
\end{figure}

The phase diagram is shown in Fig.~4.
For weak interactions, there is a SC--Fermi Insulator
(SC--FI)
transition with disorder as seen in one class of experiments.\cite{VALLES}
For a stronger, but nevertheless intermediate coupling, the phase
diagram already exhibits the SC--Bose Insulator (SC--BI) transition seen in
another class of experiments.\cite{PAALANEN}   
From Eq.~4 it follows that for small $t$ the pseudogap in
$N(\omega)$ opens even for very small $|U|/V$ at large
$\beta$, indicating that the FI--BI crossover
has the trajectory shown.\cite{USUNPUB}
Theoretically, a universal resistivity is
predicted by a bosonic treatment of the transition.\cite{FISHER}  
It has been suggested that the presence of fermions causes non-universal
properties at the SC--I transition.\cite{YAZDANI}
It is possible that the attractive Hubbard Model will exhibit universal
resistivity only in the Bose--Insulator regime,
though considerable further work needs to be done to
settle this issue.

The interplay of finite--size effects and the development of a gap
in $N(\omega)$ is a subtle issue.  As described above, in the 
case of the half--filled repulsive Hubbard model
at $U=4t$, an antiferromagnetic gap in 
$N(\omega)$ at finite temperature disappears when the lattice size
is increased beyond the correlation length of antiferromagnetic
order.  We have studied this issue carefully by further
evaluation of $N(\omega)$ on a range of lattices, and conclude that
the gap we observe is a robust feature and not an artifact of finite lattice
size.  A crucial difference here from the case of 
the half--filled repulsive model is that the antiferromagnetic transition 
takes place only at $T=0$, whereas the superconducting transition
off half--filling in the attractive model takes place at 
finite temperature.

\begin{figure}
\hspace*{-0.2in}
\psfig{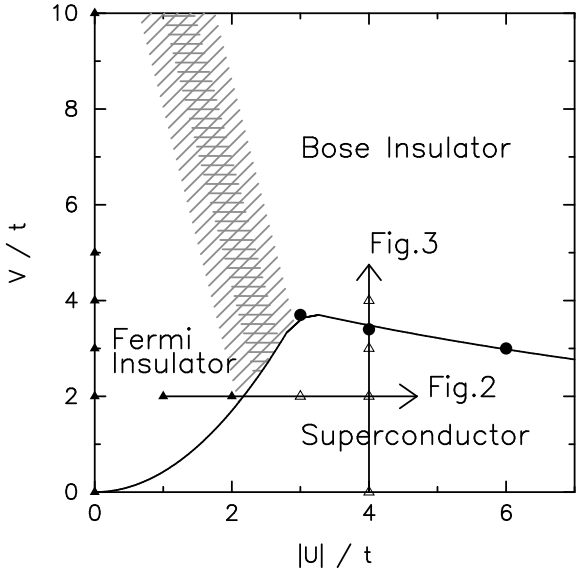}
\bigskip
\caption{             
Phase diagram as a function
of disorder $V/t$ and interaction strength $|U|/t$.  The solid
circles indicate the location of the entry into the
superconducting phase as determined by our simulations.  The remainder
of the solid line indicates an estimate of the location of the boundary
of the superconducting region.  The arrows correspond to
Figs.~2 and 3, open triangles indicate gapped regions while
closed triangles indicate no gap in $N(\omega)$.
In the superconducting regime the pair correlations have long range order and
there is a gap in $N(\omega)$.  In the insulating phases there is
no pair long range order.  
The hatched region demarks schematically the cross-over between the
Bose and Fermi Insulators.
}
\end{figure}

In conclusion, we have computed the density
of states $N(\omega)$ in a disordered interacting fermion model
with a combination of the QMC and ME methods.  The disorder--induced 
broadening expected in the weak--coupling BCS limit is
already absent by the time $|U|=4t$, half the single particle
bandwidth $W=8t$.  Disorder
closes the gap in the density of states for weak interactions but for
stronger interactions has little effect on the gap well beyond the
point where pair coherence vanishes and the system is insulating.
At the coupling where a metallic state has been previously identified,
the density of states suggests bosonic excitations on either side of
the SC--I transition.

We gratefully acknowledge support from the SDSC and the NSF
under grant No.~DMR--9528535, as well as useful
discussions with N.~Trivedi and M.~Ulmke.


\begin{references}
\bibitem{ERRORS}
This treatment of errors and qualifying both the data and the results
of a ME analytic continuation 
in the non-disordered case are discussed at length
in Ref.~\cite{GUBERNATISandJARRELL}.
\end{references}
\end{document}